\documentclass[aps,prl,twocolumn,nofootinbib,reprint]{revtex4-2}
\usepackage{amsmath,amssymb, amsfonts, bbm}
\usepackage{graphics,epsfig,color,subfigure}
\usepackage{hyperref}
\usepackage{verbatim}
\usepackage{eufrak}
\usepackage{mathtools}
\usepackage{amsthm}
\usepackage{graphicx}
\usepackage{color}

\renewcommand{\thanks}[1]{\footnote{#1}}
\newcommand{\doiurl}[2]{{\href{http://dx.doi.org/#2}{#1}}}

\newcommand{\bea}{\begin{eqnarray}}
\newcommand{\eea}{\end{eqnarray}}
\newcommand{\be}{\begin{eqnarray}}
\newcommand{\ee}{\end{eqnarray}}

\def\ie{\begin{equation}\begin{aligned}}
\def\fe{\end{aligned}\end{equation}}

\def\ie{\begin{equation}\begin{aligned}}
\def\fe{\end{aligned}\end{equation}}

\def\cN{{\cal N}}

\hypersetup{
    colorlinks=true,       
    linkcolor=red,          
    citecolor=blue,        
    filecolor=magenta,      
    urlcolor=blue           
}

\begin{document}

\title{A note on  't Hooft-line defect integrated correlators\\
 in $\mathcal{N}=4$ supersymmetric Yang--Mills theory}
\author{Daniele Dorigoni \vspace{0.2cm}}
\affiliation{ Centre for Particle Theory \& Department of Mathematical Sciences,  Durham University, Durham DH1 3LE, UK }
\affiliation{ 
Max-Planck-Institut f\"ur Gravitationsphysik (Albert-Einstein-Institut), 
am M\"uhlenberg 1, Potsdam, 14476, Germany}
\begin{abstract}
We derive the perturbative expansion of a particular integrated correlator of two superconformal primary operators in the stress tensor multiplet of $\mathcal{N}=4$ supersymmetric $SU(N)$ Yang--Mills theory in the presence of a half-BPS 't Hooft-line defect. 
The calculation is based on a recently derived expression for this physical observable in terms of a two-dimensional lattice sum with manifest automorphic properties under the electromagnetic duality group. 
When the gauge group is $SU(2)$, this analysis matches with the presented supersymmetric localisation approach while for higher-rank gauge groups, where no alternative formulation is available, the methods introduced prove to be crucial in obtaining the perturbative expansion of integrated correlators for 't Hooft-line defects.

\end{abstract}

\maketitle

\section{Introduction} 
Four dimensional $\cN=4$  supersymmetric Yang--Mills (SYM) theory  \cite{Brink:1976bc} is possibly the most promising non-trivial quantum field theory for which an exact solution may be within reach.  It is the unique four-dimensional conformal invariant field theory with maximal supersymmetry, and via the celebrated AdS/CFT correspondence it is holographically dual to type IIB superstring theory in $AdS_5\times S^5$. 
Due to its highly symmetric nature, $\mathcal{N}=4$ SYM  is one of the few theories for which many of its observables can be determined analytically. 

A crucial ingredient for this task is electromagnetic duality, usually called $S$-duality \cite{Montonen:1977sn, Witten:1978mh, Osborn:1979tq, Goddard:1976qe}, which connects in a non-trivial way  $\mathcal{N}=4$ SYM at weak coupling with the strong coupling regime of the same theory.
Following usual conventions, we denote the complex Yang--Mills coupling constant by 
\begin{equation}
\tau = \tau_1+i \tau_2 := \frac{\theta}{2\pi} + i  \frac{4\pi   }{g_{_{{\rm YM}}}^2}\,,
\end{equation}
 with $\theta$ the topological theta angle and $g_{_{{\rm YM}}}$ the Yang--Mills gauge coupling constant. A fascinating consequence of $S$-duality is that different physical observables may be related to one another upon an ${\rm SL}(2, \mathbb{Z})$ transformation of the coupling $\tau$  via
 \begin{align}
 \tau \rightarrow \tau' &\label{eq:GNOtau}= \gamma\cdot \tau \coloneqq \frac{a\tau+b}{c\tau+d}\,, 
 \end{align}
 with $ \gamma=\left(\begin{smallmatrix} a& b \\ c & d \end{smallmatrix}\right) \in {\rm SL}(2, \mathbb{Z})$.
 
Recent works have shown that $S$-duality predictions provide a crucial tool in deriving exact expressions, as non-trivial functions of the coupling constant $\tau$, for a class of observables known as \textit{integrated correlators}. In particular, starting from a matrix model formulation~\cite{Binder:2019jwn,Chester:2020dja,  Chester:2019jas, Chester:2020vyz} introduced correlation functions of four superconformal primary operators in the stress tensor multiplet integrated over some specific measures for the space-time insertion points.    
 In many cases, it has been shown that these physical observables can be expressed in terms of modular forms with non-holomorphic dependence on the coupling $\tau$, providing explicit examples of $S$-duality in $\mathcal{N}=4$ SYM~\cite{Chester:2019jas, Chester:2020vyz, Dorigoni:2021guq,Dorigoni:2021bvj,Dorigoni:2021rdo, Dorigoni:2022cua, Collier:2022emf, Dorigoni:2022zcr, Alday:2023pet, Dorigoni:2023ezg, Paul:2022piq, Paul:2023rka, Brown:2023cpz}. 
 
Importantly, we note that the complete spectrum of the theory does contain non-local defect operators as well. Correlation functions involving extended operators are extremely difficult to compute even just semi-classically, yet they are of crucial importance in understanding the theory at the non-perturbative level. The best understood example of non-local operators in $\mathcal{N}=4$ SYM are line defects, which in the holographic dual type IIB superstring theory correspond to extended strings.
 
 Of particular significance to this paper are correlation functions of two superconformal primaries in the stress tensor multiplet in the presence of a half-BPS line defect,  denoted by $\mathbb{L}_{(p,q)}$, in the fundamental representation of the gauge group $SU(N)$ and labelled by electromagnetic charges $(p,q)$ with $p$ and $q$ coprime integers. The integrated line defect correlator here considered has been introduced in \cite{Pufu:2023vwo}, and it is schematically given by
\begin{equation}
\begin{aligned} \label{eq:DD}
 \mathcal{I}_{\mathbb{L}, N} (p,q; \tau) &=   \int   \langle\,     \mathcal{O}_2(x_1) \mathcal{O}_2(x_2) \,\mathbb{L}_{(p,q)}  \rangle_{\rm c}\, {\rm d}\mu(x_i) \, ,
\end{aligned}
\end{equation}
where $\mathcal{O}_2$ is the dimension-two half-BPS superconformal primary operator in the stress tensor multiplet and the subscript ${\rm c}$ denotes the connected part of the correlator. 
The explicit form of the integration measure ${\rm d}\mu(x_i)$ appearing in \eqref{eq:DD} as well as the precise form for the correlator can be found in \cite{Billo:2023ncz,Billo:2024kri,Dempsey:2024vkf}, importantly we stress that this measure is dictated entirely by supersymmetry.

The Wilson-line defect, denoted  as $\mathbb{W} = \mathbb{L}_{(1,0)}$, can be described via a path-ordered exponential of local fields along the line supporting the defect and amounts to inserting in the path-integral the world-line of a point-like electric particle transforming in the fundamental representation of the gauge group $SU(N)$. Similarly, the ’t Hooft-line defect, $\mathbb{T} = \mathbb{L}_{(0,1)}$, is an example of a disorder operator and its path-integral definition \cite{tHooft:1977nqb} involves specifying a certain singular gauge transformation around a path that links non-trivially the line supporting the 't Hooft defect, thus effectively inserting in the path-integral a magnetic monopole which creates a magnetic flux tube along the loop.
At large $N$ and fixed $\tau$, these integrated line defects become crucial \cite{Pufu:2023vwo} in understanding scattering amplitudes of two gravitons from
extended $(p, q)$-strings in the dual type IIB superstring theory.

In \cite{Pufu:2023vwo}, the half-BPS Wilson-line defect integrated correlator is obtained indirectly from the well-known matrix model formulation for 
the expectation value of the half-BPS fundamental Wilson loop  $\mathcal{N}=2^*$ SYM on $S^4$, denoted by $\langle \mathbb{W} \rangle^{SU(N)}_{\mathcal{N}=2^*}$ and   determined by Pestun using supersymmetric localisation \cite{Pestun:2007rz}.
$\mathcal{N}=2^*$ SYM is a massive deformation, with mass parameter $m$, of the superconformal $\mathcal{N}= 4$ SYM theory and as shown in \cite{Pufu:2023vwo} the $\mathcal{N}=4$ SYM Wilson-line defect integrated correlator
$\mathcal{I}_{\mathbb{W}, N} ( \tau)$ introduced in \eqref{eq:DD} is then related to the  expectation value of the $SU(N)$ Wilson-line defect in $\mathcal{N}=2^*$ SYM as follows,
\begin{equation}\label{eq:IntroW}
\begin{aligned}
\mathcal{I}_{\mathbb{W}, N} ( \tau)= \Big[\partial_m^2 \log   \langle\,   \mathbb{W}   \,  \rangle^{SU(N)}_{\mathcal{N}=2^*}(m,\tau)\Big]_{m=0}\, . 
\end{aligned}
\end{equation}

Under $\mathcal{N}=4$ electromagnetic duality, line defects do transform non-trivially: for the theory with coupling constant $\tau'=\gamma \cdot \tau$ given in \eqref{eq:GNOtau}, the line defect $\mathbb{L}_{(p,q)}$ is mapped into a defect with charges $(p',q')$ given by
 \begin{align}
\mathbb{L}_{(p,q)} \rightarrow  \mathbb{L}_{(p'\!,q')} \, ,  \quad {\rm with}\quad  \, (\begin{smallmatrix} p'\!,\,q' \!\end{smallmatrix})= (\begin{smallmatrix} p\phantom{'}\!,\,q \end{smallmatrix}) \left(\begin{smallmatrix} a& -c \\ -b & d \end{smallmatrix}\right) \, . \label{eq:GNOcharges}
 \end{align}
 This implies that correlation functions in the presence of a line defect operator such as \eqref{eq:DD} must obey the following transformation properties, 
 \begin{equation}
\begin{aligned} \label{eq:autoformIntro}
 \mathcal{I}_{\mathbb{L}, N} (p,q;  \tau) =  \mathcal{I}_{\mathbb{L}, N} (p'\!,q';  \tau')\, ,
\end{aligned}
 \end{equation}
 valid for all $\gamma\in {\rm SL}(2,\mathbb{Z})$ when the coupling constant $\tau' = \gamma\cdot \tau$ and the charges $(p',q')$ have been transformed accordingly to \eqref{eq:GNOtau} and \eqref{eq:GNOcharges}.
 
Recently \cite{Dorigoni:2024vrb} combined the property \eqref{eq:autoformIntro} with an explicit matrix model computation to conjecture the lattice sum integral representation of $ \mathcal{I}_{\mathbb{L},N}(p,q;\tau)$ valid for any $N$ and any defect-charges  $(p,q)$,
\begin{align}
&\label{eq:LatticeSumRes2} \mathcal{I}_{\mathbb{L},N}(p,q;\tau)=  \frac{N}{L^1_{N-1}( -\frac{\pi}{\tau'_2})}   \times \\
&\notag\! \sum_{(n,m)\in \mathbb{Z}^2} \! \int_0^\infty  \!\!\! e^{- t_2 \pi \frac{|n\tau+m|^2}{\tau_2}-t_3 \pi \tau_2'   (np-mq)^2}  \widetilde{\mathcal{B}}_N(\frac{ \tau_2'}{\pi}, t_2,t_3) {\rm d}^2t,
\end{align}
where we defined $\tau_2' \coloneqq \tau_2/|q\tau+p|^2$ and $\widetilde{\mathcal{B}}_N(y, t_2,t_3)$ is a function of the three real variables $y, t_2, t_3$ satisfying the inversion transformation and integral identity
\begin{align}
&{\widetilde{\mathcal{B}}}_N(y, t_2,t_3) \label{eq:inv1}= \frac{{\widetilde{\mathcal{B}}}_N\big(y, \frac{t_2}{t_2(t_2+t_3)}, \frac{t_3}{t_2(t_2+t_3)} \big)}{[t_2(t_2+t_3)]^{\frac{3}{2}}} \,,\\
&\!\!\!\! \int_0^\infty t_2^{-\frac{1}{2}} \,{\widetilde{\mathcal{B}}}_N(y, t_2,t_3)\, {\rm d}^2 t \label{eq:inv2}= 0\,, \quad \forall \,y>0\,.
\end{align}
 The overall factor in \eqref{eq:LatticeSumRes2} given in terms of a generalised Laguerre polynomial, $L^1_{N-1}(x)$, arises from the normalisation of the integrated correlator~\eqref{eq:DD} and it is related to the vacuum expectation value of the line defect operator.

In this work, we focus our attention to the case of the 't Hooft-line defect integrated correlator.
Firstly, similar to the integrated Wilson-line defect definition \eqref{eq:IntroW} we use the matrix model formulation proposed in \cite{Gomis:2011pf} for the expectation value of an 't Hooft-line defect in  $\mathcal{N}=2^*$ SYM with gauge group $SU(2)$, to compute the perturbative expansion of the $\mathcal{N}=4$ 't Hooft-line defect integrated correlator as $g_{_{\rm YM}}\to0$, or equivalently for $\tau_2\gg1$.
We show that this calculation matches identically with the perturbative expansion of the lattice sum \eqref{eq:LatticeSumRes2} specialised to the case of an 't Hooft-line defect $(p,q)=(0,1)$ and  $N=2$. We repeat our analysis for the case of higher-rank gauge groups, in particular for $SU(3)$, for which no alternative method is available to compute the perturbative expansion of the 't Hooft-line defect integrated correlator. 
\\

\section{ $SU(2)$ Matrix model formulation} 
 
Thanks to supersymmetric localisation, \cite{Gomis:2011pf} provided a matrix model integral representation for the expectation value of a half-BPS fundamental 't Hooft-line defect in $\mathcal{N}=2^*$ SYM positioned on the equator of $S^4$ which for the case of an $SU(2)$ gauge group takes the form
\begin{align}
&\label{eq:Tvev2} \langle \mathbb{T} \rangle^{SU(2)}_{\mathcal{N}=2^*} (m,\tau)= [Z_2(m,\tau)]^{-1}\times  \\
& \notag  \int_{-\infty}^{\infty} \vert \,Z^{\rm cl}(\tau,a) Z^{\rm pert}(m,a) Z^{\rm inst}(\tau,m,a) |^2 \,Z^{\rm eq}(m,a) \,{\rm d} a\,.
\end{align}
The integral runs over the one-dimensional Cartan subalgebra of $SU(2)$ parametrised by $a$ and the normalisation factor $Z_2(m,\tau)$ denotes Pestun \cite{Pestun:2007rz} partition function for $\mathcal{N}=2^*$ SYM on $S^4$ and gauge group $SU(2)$ .
The classical action contribution can be written as
\begin{equation}\label{eq:Zcl}
|Z^{\rm cl}(\tau,a)|^2 = \exp\Big[ \frac{ \pi |\tau|^2}{4\tau_2 } - 4 \pi \tau_2 \Big(a+\frac{\tau_1}{4\tau_2}\Big)^2\Big]\,.
\end{equation}
The term $Z^{\rm pert}(m,a)$ encodes the one-loop determinant fluctuations while $Z^{\rm inst}(\tau,m,a)$ is expressible in terms of Nekrasov partition function \cite{Nekrasov:2002qd} and describes the contributions from instantons and anti-instantons localised at the poles of $S^4$. Since we are interested in the purely perturbative sector in the limit $\tau_2\gg1$ we set $Z^{\rm inst}(\tau,m,a) \to 1$ in what follows.
Finally, the factor $Z^{\rm eq}(m,a)$ encodes all contributions to the path-integral which have support precisely on the equator of the $S^4$ where the 't Hooft-line defect has been inserted. Besides an important perturbative part, these equatorial contributions contain crucial non-perturbative effects due to monopole bubbling where we need to include in the path-integral smooth monopoles configurations which screen the magnetic charge of the 't Hooft-line defect. 
In  \cite{Gomis:2011pf} an exact expression for the complete $Z^{\rm eq}(m,a)$ is provided only in the case of $\mathcal{N}=2^*$ SYM with gauge group $SU(2)$, which for the present case of the minimal fundamental 't Hooft-line defect it amounts to
\begin{equation}
Z^{\rm eq}(m,a) \coloneqq 2\frac{ \sqrt{\cosh[\pi(2a+m)] \cosh[\pi(2a-m)]}}{\cosh(2\pi a)}\,.
\end{equation}

We now follow the definition \eqref{eq:IntroW} and consider the integrated 't Hooft-line defect correlator in $\mathcal{N}=4$ SYM 
\begin{equation}\label{eq:IntroT}
\begin{aligned}
\mathcal{I}_{\mathbb{T}, 2} ( \tau)= \Big[\partial_m^2 \log   \langle\,   \mathbb{T}   \,  \rangle^{SU(2)}_{\mathcal{N}=2^*}(m,\tau)\Big]_{m=0}\, .
\end{aligned}
\end{equation}
To compute the perturbative sector of \eqref{eq:IntroT} as $\tau_2\gg1$, we use the matrix model expression \eqref{eq:Tvev} and discard all exponentially suppressed corrections, thus arriving at
\begin{equation}\label{eq:Tint}
\mathcal{I}^{\rm pert}_{\mathbb{T}, 2} ( \tau) = \frac{ \langle \mathbb{T} \mathcal{O}_2\mathcal{O}_2\rangle^{\rm pert} -\langle \mathbb{T}\rangle \langle \mathcal{O}_2 \mathcal{O}_2\rangle^{\rm pert}  }{\langle \mathbb{T}\rangle }\,.
\end{equation}
For the perturbative defect two-point function we have
\begin{align}
&\label{eq:IT2}\!\!\! \langle \mathbb{T} \mathcal{O}_2\mathcal{O}_2\rangle^{\rm pert}(\tau) \coloneqq  [Z_2(0,\tau)]^{-1}    \int_{-\infty}^{\infty} \vert \,Z^{\rm cl}(\tau,a)\vert^2 \\
&\notag \!\!\!\! \times\!\!  \Big[ 2\Big(\partial_m^2 Z^{\rm pert}(m,a)\vert_{m=0} +{\rm c.c.}\Big) + \partial_m^2Z^{\rm eq}(m,a) \vert_{m=0} \Big]  {\rm d} a,
\end{align}
with ${\rm c.c.}$ the complex conjugate term, as well as
\begin{equation}
\langle \mathcal{O}_2 \mathcal{O}_2\rangle^{\rm pert} (\tau) \coloneqq \partial_m^2 Z_2(m,\tau)\vert_{m=0}\,.
\end{equation}
The denominator appearing in \eqref{eq:Tint} is the aforementioned expectation value of the half-BPS fundamental 't Hooft-line defect in $\mathcal{N}=4$ SYM which is known for general $SU(N)$ gauge group from the Wilson-line result \cite{Drukker:2000rr}, 
\begin{equation}\label{eq:Tvev}
\langle \mathbb{T}\rangle(\tau) = \frac{1}{N} e^{\frac{N-1}{N}  \frac{\pi |\tau|^2}{2 \tau_2 } } L_{N-1}^1 \big( - \frac{\pi |\tau|^2}{\tau_2} \big)\,.
\end{equation}

The disconnected term in \eqref{eq:Tint} is well-understood and it is essentially the perturbative contribution to the integrated four-point correlator evaluated in \cite{Chester:2019pvm}. To compute the weak coupling expansion of \eqref{eq:Tint} we also need
\begin{align}
&\partial_m^2 Z^{\rm pert}(m,a)\vert_{m=0} = \\
&\notag  \big(4a^2+\frac{1}{4}\big) \times\big[2  \psi(2i a + \tfrac{1}{2})+4 i a \psi'(2ia + \tfrac{1}{2})+2\gamma\big] \,,
\end{align}
where the overall factor comes from the $SU(2)$ Vandermonde determinant shifted by the 't Hooft-line magnetic charge,
 $\psi(x)$ denotes the polygamma function and $\gamma$ is the Euler-Mascheroni constant.
The integral in \eqref{eq:IT2} can be easily performed by shifting integration variables $a\to \tilde{a} = a+ \tau_1/(4\tau_2)$ and then expand everything but the classical action for $\tilde{a}\to0$, thus reducing everything to a collection of gaussian integrals. Importantly, we notice that the exponential contribution coming from the classical action \eqref{eq:Zcl} cancels exactly against the same contribution appearing at denominator and coming from the  vacuum expectation value \eqref{eq:Tvev} with $N=2$.

We are then left with the perturbative expansion for the $SU(2)$ 't Hooft-line defect integrated correlator
\begin{align}\label{eq:Tint2Pert}
&\mathcal{I}^{\rm pert}_{\mathbb{T}, 2} ( \tau)  {=}  \frac{2}{L^1_{1}( -\frac{\pi | \tau|^2}{\tau_2})}  \sum_{\ell=0}^\infty c_\ell(\tau_1 ) (\pi \tau_2)^{1-\ell} \,,
\end{align}
where at the first few orders we find
\begin{align}
c_0(\tau_1) &\notag = - 8   \log (2)+\pi ^2\,,\\
c_1(\tau_1) &\notag = -16 \log (2) +2 \pi ^2+24 \zeta (3)-\frac{\pi ^4}{2} \,,\\
c_2(\tau_1) &\label{eq:coeff} =216 \zeta (3)-3 \pi ^4 -390 \zeta (5)+\frac{\pi ^6}{2}\\
&\notag \phantom{=} + (\pi\tau_1)^2 \big[-8 \log (2)+\pi ^2+ 21 \zeta (3)-\frac{\pi ^4}{4}\big]\,,\\
c_3(\tau_1) &\notag= -4500 \zeta (5)+5 \pi ^6+6300 \zeta (7)-\frac{17 \pi ^8}{24}\\
&\notag \phantom{=}+(\pi \tau_1)^2 \big[ 234 \zeta (3)-3 \pi ^4-465 \zeta (5)+\frac{\pi ^6}{2}\big]\,.
\end{align}
We notice that $c_\ell(\tau_1)$ is a polynomial of degree $\lfloor \ell /2 \rfloor$ in $\tau_1^2$, hence in particular the perturbative expansion of  the 't Hooft-line defect integrated correlator does in fact depend from the topological theta angle $\theta = 2\pi\tau_1$. This is a consequence of the Witten effect \cite{Witten:1978mh} telling us that a translation of the $\theta$ angle, $\tau \to \tau+n$, does in fact induce a modification of the electric charge $p$ for a line defect $\mathbb{L}_{p,q}$ with a non-vanishing magnetic charge $q\neq0$, as we can see directly from the transformation properties \eqref{eq:GNOcharges}-\eqref{eq:autoformIntro}.
From a number theoretical point of view the structure of the coefficients $c_\ell(\tau_1)$ is also very interesting: at order $\ell$ the numbers appearing in $c_\ell(\tau_1)$ have trascendentality between $2\ell-1$ and $2\ell+2$ where we assign standard trascendentality $[\zeta(n)]=n$, $[\log(2)]=1$ and $[( \tau_1 \pi)]=1$ with $[r]=0$ for $r\in \mathbb{Q}$.

We are now going to show that the very same perturbative expansion \eqref{eq:Tint2Pert} can be obtained from the lattice sum representation \eqref{eq:LatticeSumRes2}, constructed in \cite{Dorigoni:2024vrb} by exploiting solely the matrix model formulation for the Wilson-line defect integrated correlator \eqref{eq:IntroW} and the crucial electromagnetic duality transformation  \eqref{eq:autoformIntro}.
\\

\section{ Electromagnetic duality predictions} 
We now consider the general lattice sum representation \eqref{eq:LatticeSumRes2} specialised to the case of the $SU(2)$ 't Hooft-line defect integrated correlator thus taking the form
\begin{align}
&\label{eq:LatticeSumRes2T} \mathcal{I}_{\mathbb{T},2}(\tau)=  \mathcal{I}_{\mathbb{L},2}(0,1;\tau) = \frac{2}{L^1_{1}( -\frac{\pi | \tau|^2}{\tau_2})} 
\widetilde{\mathcal{I}}_{\mathbb{T},2}(\tau)\,,
\end{align}
where we defined the `reduced' integrated correlator
\begin{align}
& \label{eq:ITtilde}\widetilde{\mathcal{I}}_{\mathbb{T},2}(\tau) =\\
&\notag  \sum_{(n,m)\in \mathbb{Z}^2}  \int_0^\infty \!\!  e^{- t_2 \pi \frac{|n\tau+m|^2}{\tau_2}-t_3 \pi\frac{ \tau_2 }{|\tau|^2}   m^2 }  \widetilde{\mathcal{B}}_2(\frac{\tau_2}{ \pi|\tau|^2}, t_2,t_3) \,{\rm d}^2t\,.
\end{align}
The function $\widetilde{\mathcal{B}}_2(y, t_2,t_3)$  has been derived in \cite{Dorigoni:2024vrb} and takes the simple form
\begin{align}
 {\widetilde{\mathcal{B}}}_2(y,t_2,t_3)  = \frac{ \exp\big[-\frac{t_3}{4 y (t_2+1)  (t_2+t_3+1)}\big] P(y,t_2,t_3)}{ y^3 (t_2+1)^{\frac{11}{2}}  (t_2+t_3+1)^{\frac{11}{2}}} \,,
\end{align}
with $P(y,t_2,t_3)$ an unenlightening polynomial in its three arguments given in equation (4.35) of the same reference.

The task at hand is extracting the perturbative expansion of \eqref{eq:ITtilde} as $\tau_2\gg1$. We proceed by first performing a Poisson resummation in the variable $m\to\hat{m}$ and then change integration variables to $( x_2,x_3) \coloneqq (t_2, (t_2+t_3)^{-1})$ so that the domain of integration becomes $x_2,x_3\geq 0 $ with $x_2 x_3\leq 1$. Note that in the new integration variables the properties \eqref{eq:inv1}-\eqref{eq:inv2} become
\begin{align}
&\label{eq:Inv} x_3^{-\frac{3}{2}} \widetilde{\mathcal{B}}_N(y,x_2,x_3) = x_2^{-\frac{3}{2}}\widetilde{\mathcal{B}}_N(y,x_3,x_2)\,,\\
&\label{eq:Inv2}\!\! \int_0^\infty \!\!\! \int_0^{\frac{1}{x_2}}\frac{1}{\sqrt{x_2 x_3}}\, x_3^{-\frac{3}{2}}\widetilde{\mathcal{B}}_N(y,x_2,x_3)\,{\rm d}^2 x = 0\,.
\end{align}

At this point we expand the integral as a power series in $\tau_1$ arriving at
\begin{align}
& \label{eq:ITtildeN}\widetilde{\mathcal{I}}_{\mathbb{T},2}(\tau) =\\
&\notag 
\!\sum_{k=0}^\infty (\pi \tau_1)^{2k}\!\!\! \! \!\sum_{(n,\hat{m})\in \mathbb{Z}^2} \! \int_0^\infty  \!\!\!\!\int_0^{\frac{1}{x_2}}  \! \!\frac{ y^{\frac{7}{2}}\,e^{-S  y } \,p_k( y,x_2,x_3,n,\hat{m})   }{[(x_2+1)(x_3+1)]^{k+\frac{11}{2}}}{\rm d}^2x,
\end{align}
where $y \coloneqq \pi \tau_2$ and we have introduced the `action'
\begin{equation}\label{eq:Seff}
S \coloneqq  x_2  n^2 + \frac{1}{4(x_2+1) }+x_3 \hat{m}^2 + \frac{1}{4(x_3+1) } - \frac{1}{4} \,.
\end{equation}
The integrand factors $p_k$ are simply polynomials in $y^{-1/2}$ and in the remaining variables, and are symmetric with respect to $(x_2,x_3,n,\hat{m})\to(x_3,x_2,\hat{m},n)$ as a consequence of \eqref{eq:Inv}.
All odd powers in $\tau_1$ vanish identically thanks to the symmetry of \eqref{eq:ITtildeN} under $n \to -n$.

To extract the asymptotic expansion of \eqref{eq:ITtildeN} for $y \gg 1$ we must analyse the action \eqref{eq:Seff}. We notice that when $x_3$  is close to the extrema of integration we have $S\geq 1/4$ as $x_3\to 0$ and $S\sim x_2 n^2 + \hat{m}^2 /x_2$ as $x_3\to 1/x_2$.
We deduce that the only perturbative contributions at large $y$ must arise from the region near $x_3 \sim 1/x_2$ when necessarily either $\hat{m}=0$ or $n=0$ or possibly both. 

Firstly, we check by direct computation that the contribution coming from $n = \hat{m}=0$ vanishes order by order in $\tau_1$.
This fact had already been appreciated in \cite{Dorigoni:2024vrb} for the different case when \eqref{eq:LatticeSumRes2} is specialised to a Wilson-line defect and the vanishing of the corresponding $n = \hat{m}=0$ term is actually equivalent to the integral identity \eqref{eq:Inv2}. However, our analysis shows that the vanishing of the $n = \hat{m}=0$ term holds more in general and in particular it holds for the present 't Hooft-line defect case. 
Given the symmetry of the integrand and of the domain of integration under the exchange $(x_2,n) \leftrightarrow (x_3 , \hat{m})$, we can then simply consider the case $\hat{m}=0, n \neq 0$ timed by an additional factor of $2$.

We then perform the integral over $x_3$ in \eqref{eq:ITtildeN} separately for each $\tau_1^{2k}$ power and find 
\begin{align}\label{eq:Ipert0}
&\widetilde{\mathcal{I}}^{\rm pert}_{\mathbb{T}, 2} ( \tau)  =\sum_{n=1}^\infty \int_0^\infty  e^{- x_2  n^2 y} \Big[  \frac{y^{\frac{3}{2}} \sqrt{x_2}\, q_1(y,x_2)}{ \sqrt{\pi } (x_2+1)^5}\\
&\notag  +
\frac{y\, e^{\frac{x_2 y}{4 x_2+4}}q_2(y,x) \,\text{erfc}\left(\frac{1}{2} \sqrt{\frac{x_2 y}{x_2+1}}\right)}{ (x_2+1)^{\frac{11}{2}}}\Big] \,{\rm d}x_2+ O(\tau_1^2)\,,
\end{align}
where we discarded exponentially suppressed corrections at large $y$ and denoted by $\text{erfc}(x)$ the complementary error function.
In the above equation we defined
\begin{align}
q_1(y,x) &= -28 - 16 x + 16 x^2 + 4 x^3 - 2y(1 - 2 x )\,,\\
q_2(y,x) &= 8 - 28 x - 72 x^2 - 28 x^3 + 8 x^4 \\
 &\notag \phantom{=}+ y(4 + 12 x - 8 x^2  - 
 16 x^3) +  y^2 (x+ 2 x^2 )\,.
\end{align}
and we note that the integral structure in \eqref{eq:Ipert0} remains identical for higher $\tau_1^{2k}$ corrections (as well as for higher rank cases $N>2$), with the only modification that at higher-order the analogues of the polynomials $q_1,q_2$ are in general polynomials also in $n^2$.

To extract the large-$y$ expansion of the term proportional to $q_1$ in  \eqref{eq:Ipert0} we simply need to expand the term in parenthesis around $x_2=0$ and then use
\begin{equation}
\sum_{n=1}^\infty \int_0^\infty  e^{- x_2 n^2 y } x_2^{s-1} \,{\rm d}x_2 = y^{-s} \zeta(2 s) \Gamma (s)\,,
\end{equation}
valid for $\mbox{Re}\,s>0$, so that higher orders in $x_2$ corresponds automatically to higher perturbative corrections in $1/y$.
The large-$y$ expansion of the first term in \eqref{eq:Ipert0} then yields
\begin{align}
&\notag \sum_{n=1}^\infty \int_0^\infty  e^{- x_2  n^2 y}   \frac{y^{\frac{3}{2}} \sqrt{x_2}\, q_1(y,x_2)}{ \sqrt{\pi } (x_2+1)^5} {\rm d}x_2\\
&\label{eq:q1} =  - y \, \zeta(3)+\left( - 14 \zeta(3)+\frac{9 \zeta(5)}{2} \right)   + O(y^{-1})\,.
\end{align}

For the second term in \eqref{eq:Ipert0}  the situation is a little bit more complicated. If we perform the same procedure as described above and expand near $x_2=0$ we obtain a power series in $1/y$ with coefficients given by convergent infinite sums over Riemann zeta values,~i.e.
\begin{align}
&\notag\sum_{n=1}^\infty \int_0^\infty  e^{- x_2  n^2 y}
\frac{y \,e^{\frac{x_2 y}{4 x_2+4}}q_2(y,x)\, \text{erfc}\left(\frac{1}{2} \sqrt{\frac{x_2 y}{x_2+1}}\right)}{2 (x_2+1)^{\frac{11}{2}}}\,{\rm d}x_2\\
&\notag = y \big[ \zeta(3)+ \sum_{n=2}^\infty  (-1)^n 2^{3-n} n \,\zeta(n) \big] + O(y^{0}) \\
&\label{eq:q2}= y   \left(-8 \log (2)+\pi ^2 +\zeta(3)\right) +O(y^0) \,.
\end{align}
Curiously, at each order in $1/y$ we see that the trascendentality of the above expansion coefficients \eqref{eq:q1}-\eqref{eq:q2} exceeds by one unit that of \eqref{eq:coeff}, e.g. at order $y$ we find a $\zeta(3)$ term. However, when we add together the two separate contributions appearing in \eqref{eq:Ipert0} we discover that these unwanted terms magically disappear and the asymptotic expansion for $y=(\pi\tau_2)\gg1$ of \eqref{eq:Ipert0} reproduces the supersymmetric localisation results \eqref{eq:Tint2Pert}-\eqref{eq:coeff}.

Thanks to the analysis of  \cite{Dorigoni:2024vrb} we can repeat an analogous story and derive the perturbative expansion of the $SU(3)$ 't Hooft-line defect integrated correlator for which we obtain:
\begin{align}
& \mathcal{I}^{\rm pert}_{\mathbb{T}, 3} ( \tau)  {=}  \frac{3}{L^1_{2}( -\frac{\pi | \tau|^2}{\tau_2})}  \Big[ (\pi \tau_2)^2 (2 \pi ^2-16 \log (2)) \\
&\notag+ (\pi \tau_2)^1 (-96 \log (2)+12 \pi ^2+72 \zeta(3)-\frac{3 \pi ^4}{2}) + O(1)\Big]\,.
\end{align}
Note that when compared to \eqref{eq:Tint2Pert} the factor $\widetilde{\mathcal{I}}^{\rm pert}_{\mathbb{T}, 3} ( \tau) $ starts at order $O(\tau_2^2)$, consequence of the fact that the denominator $L^1_{2}(-y)$ is now a degree $2$ polynomial while $L^1_{1}(-y)$ which appears in \eqref{eq:Tint2Pert} has degree one.  With our method we can furthermore compute the $\tau_1$ terms appearing in the perturbative expansion of the 't Hooft-line defect which for $\widetilde{\mathcal{I}}^{\rm pert}_{\mathbb{T}, 3} ( \tau) $ first appear at order $O(y^0)$. 
We stress that the $SU(2)$ matrix model formulation \eqref{eq:IT2} is not explicitly known for 't Hooft-line defect correlator with higher-rank gauge groups due to the complicated nature of monopole bubbling effects in \cite{Gomis:2011pf}. Nonetheless, our results provide a concrete way to access the perturbative expansion of general integrated 't Hooft-line defect correlators.

\section{Conclusions}

In this letter we have confirmed that the lattice sum representation \eqref{eq:LatticeSumRes2}, constructed in  \cite{Dorigoni:2024vrb} by using only the automorphic properties under electromagnetic ${\rm SL}(2,\mathbb{Z})$ duality of the $\mathcal{N}=4$ SYM integrated Wilson-line defect operator, does in fact reproduce identically \footnote{As a further check beyond perturbation theory, \cite{Dorigoni:2024vrb} evaluated numerically the matrix model integral for the $SU(2)$ 't Hooft-line defect integrated correlator and showed agreement with the lattice sum formulation within the numerical precision used. } the perturbative expansion of the $SU(2)$ integrated 't Hooft-line defect correlator derived from the $\mathcal{N}=2^*$ SYM supersymmetric localisation results of \cite{Gomis:2011pf}.
Furthermore, we have shown how to extract from the proposal of  \cite{Dorigoni:2024vrb} the perturbative expansion of integrated 't Hooft-line defect correlators for higher-rank gauge groups for which no alternative method is presently available, presenting here explicitly the $SU(3)$ case.

The methods proposed bypass entirely the challenging, if not impossible task of computing the un-integrated 't Hooft-line defect two-point function by expanding the path-integral around a monopole background and then integrate over the insertion points as in \eqref{eq:DD}.
Very little is known about the semiclassical expansion of 't Hooft-line correlation functions, see e.g. \cite{Kristjansen:2023ysz}, however our perturbative data do provide for precious checks, via \eqref{eq:DD}, against future un-integrated 't Hooft line defects results, similar to what has been done for the integrated correlators of four superconformal primaries of the stress tensor multiplet in \cite{Wen:2022oky,Zhang:2024ypu}.

\textit{Acknowledgments.} DD thanks Zhihao Duan, Congkao Wen and Haitian Xie for related discussions and Congkao Wen for comments on the draft. DD  is  indebted to the Albert Einstein Institute and in particular to Axel Kleinschmidt, Hermann Nicolai and Stefan Theisen
for the hospitality and financial support.




\end{document}